\documentclass[preprint,prX,
floatfix,superscriptaddress]{revtex4-1}
\usepackage{bm,amsmath,amssymb,mathrsfs}
\usepackage{cancel,epsfig,graphicx}
\begin{document}
\title{Parity nonconserving proton-proton elastic
scattering}
\author{T.~M.~Partanen}
\email{tero.partanen@helsinki.fi}
\author{J.~A.~Niskanen}
\email{jouni.niskanen@helsinki.fi}
\affiliation{ Department of Physical Sciences, P.~O.
~Box~64, FIN-00014 University of Helsinki, Finland}
\author{M.~J.~Iqbal}
\email{iqbal@phas.ubc.ca}
\affiliation{ Department of Physics and Astronomy,
University of British Columbia, Vancouver, BC, Canada,
V6T 1Z1}
%
%
\begin{abstract}
The parity nonconserving longitudinal analyzing power
$\bar{A}_L$ is calculated in elastic $\vec{p}p$ scattering
at the energies below the approximate inelastic region
$T_{\rm lab} = 350$ MeV. The short-ranged heavy meson $\rho$
and $\omega$ exchanges as well as the longer-ranged $2\pi$
exchanges are considered as the mediators of the parity
nonconserving interactions. The DDH "best" coupling values
are used as the parity nonconserving meson-$NN$ couplings.
Also three different parity nonconserving two-pion exchange
potentials by various authors are compared.
\end{abstract}
\maketitle
\section{Introduction}
Weak interaction is distinct in the leptonic, semileptonic,
and strangeness nonconserving hadronic processes. However,
it is not so clear-cut in the strangeness conserving
hadronic sector due to its diminutive strength against
that of incessantly present strong interaction.
Nevertheless, the parity nonconserving (PNC) weak
interaction is unique in the sense that it sorts out
different helicity states unlike any other interaction. For
this particular reason, it can, in principle, be extracted
under those overwhelming and unfailingly parity conserving
(PC) strong and electromagnetic interactions.

Even though a direct heavy $Z^0$ or $W^\pm$ boson exchange
is highly improbable over the internuclear distances, it is
feasible between the nucleon and virtual meson.
Consequently, the PNC $NN$ interactions may be parametrized
by weak meson-$NN$ coupling constants modelled in terms of
quarks and intermediate bosons. Traditionally the
PNC $NN$ calculations have relied largely on the single
meson exchange picture, based on the DDH potential
\cite{ddh} in which the PNC $NN$ interactions are due to
$\pi^\pm$, $\rho$, and $\omega$ exchanges. Nowadays at very
low energies, the calculations are preferably done in the
framework of the model-independent effective field theories
(EFT). However, all these models are parameterized by about
half a dozen weak meson-$NN$ couplings (see {\it e.g.}
Refs. \cite{zhuetal,rammus}), which are, even today,
insufficiently known despite all the experimental and
theoretical efforts.

Due to the fact the PNC interactions treat unequally
different helicity states, the PNC nucleon-nucleon ($NN$)
experiments are inherently based on the spin control of the
particle. Probably the cleanest observable, in the sense
that it is nearly a $100\%$ pion exchange dominated, arises
from the radiative PNC reaction
$\vec{n}p\rightarrow\gamma d$ at threshold. The ongoing
NPDGamma experiment \cite{npdgamma} aims to determine the
weak $\pi NN$-coupling $h_\pi^{(1)}$ by measuring the
$\gamma$-asymmetry of this reaction, with such an accuracy
that should elucidate the correctness of the most preferred
value of the $h_\pi^{(1)}=4.6\times 10^{-7}$ suggested by
DDH. Instead, the PNC $\vec{\gamma}d\leftrightarrow np$
reactions at threshold would lead only to nonpionic
exchange effects, despite of small exchange currents and
$\Delta$-effects \cite{part}.
In any case, when it comes to elastic PNC $\vec{p}p$
scattering, it is generally believed (based on the simple
single meson exchange picture) that the pion does not
contribute to it due to the lack of $\pi^0$-exchange. This
is because in general the PNC neutral spinless meson
exchange, ${\it e.g.}$ $\pi^0$, is forbidden by the
simultaneous violations of the P and CP symmetries
\cite{barton}.
However, not only the fact that the strong and weak (DDH)
pion couplings are sizable, but also that the pions are
nearly six times lighter than heavy mesons, it seems
reasonable to assume that the longest-range and possibly
the leading effects are nonetheless due to pion exchange
(in this particular case, induced by the two charged pions).

For the measurement of the PNC $\vec{p}p$ longitudinal
analyzing power $\bar{A}_L$, there exist three precision
experimental data points: Bonn at 13.6 MeV
$(-0.93\pm0.21)\times 10^{-7}$ \cite{bonn}, PSI at 45 MeV
$(-1.50\pm0.22)\times 10^{-7}$ \cite{sin}, and TRIUMF at
221.3 MeV  $(0.84\pm0.29)\times 10^{-7}$ \cite{triumf}.
The Bonn and PSI experiments are low energy scattering
experiments, where the contribution to the $\bar{A}_L$
arises only from the lowest ${}^1S_0-{}^3P_0$ transition.
The TRIUMF experiment, on the contrary, is a transmission
experiment with the energy chosen so that the contribution
arises merely from the ${}^3P_2-{}^1D_2$ and
${}^1D_2-{}^3F_2$ transitions. At the energy of the TRIUMF
experiment, the ${}^1S_0$ and ${}^3P_0$ phases
serendipitously cancel out due to strong interaction
interference from which follows that the $J=0$ transition
goes to zero, while the $J=4$ and higher ones still remain
insignificant.
What is more, for $J=2$, the local and nonlocal
contributions of the
$\omega$ exchange mostly cancel out because of a small
isoscalar anomalous magnetic moment $\chi_\omega$.
In contrast for the $\rho$ exchange, the local
contributions dominate over the nonlocal ones because of
a large isovector anomalous magnetic moment $\chi_\rho$.

Assuming that the $J=2$ mixing arises from the $\rho$
exchange, the central goal of the TRIUMF experiment was
to determine the weak $\rho pp$-coupling $h_\rho^{pp}=
h_\rho^{(0)}+h_\rho^{(1)}+h_\rho^{(2)}/\sqrt{6}$ whereas
the lower-energy experiments Bonn and PSI determined the
$h_\rho^{pp}+h_\omega^{pp}$, where
$h_\omega^{pp}=h_\omega^{(0)}+h_\omega^{(1)}$. In these
experiments the reasoning was built on the DDH potential.
However, already the work \cite{coupled} including the
effect of intermediate $N\Delta(1232)$ states via $\rho$
exchanges in the coupled channels showed that the simplest
and most straightforward interpretation of the TRIUMF
experiment might not be enough. The $\Delta$ effect was
significant enough to suggest that the coupling could
rather be effective involving $\rho$ exchange both in $NN$
and $NN \leftrightarrow N\Delta$ transitions.
In our later work Ref. \cite{our1} on PNC $\vec{p}p$
elastic scattering, we looked at the effects of the
$N\Delta$-channels in the coupled-channels formalism
as well as the effects of the two-pion exchange (TPE). The
effects were again found significant and cast doubt on the
aforementioned $h_\rho^{pp}$-coupling and whether its value
is straightforwardly proportional to the TRIUMF data point.
The preceding works Refs. \cite{liuhydes,liu} on the
reaction in question take into account the TPE,
of which the former investigates it as a part of the
short-ranged $\rho$ meson exchange and the latter
considers it in the framework of the EFT.
%

As for the present work, we should stress that the purpose
of this paper is no more than to emphasize the importance of
the TPE in the PNC $\vec{p}p$ elastic scattering, which
should be clear from the related model dependencies. Even
though the TPE is far more complicated than the single
meson exchange, it should not be ignored in this particular
case due to its considerable strength and range.
As shown in Ref. \cite{our1}, another possibly noteworthy
contribution arises from the $\Delta$-resonance even at low
energies, but it is not taken into consideration here in its
fullest form because of the large uncertainties related to
the meson-$N\Delta$ couplings especially in the weak sector. The $\Delta$ is taken into account only to the
extent it appears in the PNC TPE potentials. Since there is
no $\pi N\Delta$ coupling related to a PNC vertex
\cite{fcdh} or it is small \cite{henley71,kais90}
(we take it as zero), then on the side of the weak
couplings, the PNC TPE effects are only proportional to
the $h_\pi^{(1)}$.
Besides the DDH, there are various calculations
\cite{fcdh,desp80,dubo86,iqbal89,kais89,henley98,lobov02,
leehj} for the the $h_\pi^{(1)}$ (ranging between $0$ and
$3.4\times10^{-7}$)
indicating a smaller value than what is the DDH "best"
recommendation.
The hope is that the NPDGamma experiment would reduce the
obscurity of this coupling constant.

This work is based on the use of the distorted-wave Born
approximation (DWBA) and the optical thorem. In the
calculations, we employ
the Reid93 potential \cite{reid93} taking into account the
lowest five parity admixed transitions, {\it i.e.} the
total angular momentum up to $J=4$.
The short-ranged contributions are taken as the results of
heavy meson $\rho$- and $\omega$- exchanges, for which we
use the DDH potential. For the long-ranged effects, we
compare three different PNC TPE potentials on the market
given in Refs. \cite{our1,desplanq,kaiser}.
Note that, besides Ref. \cite{kaiser}, there exists also
another chiral perturbation theory (ChPT) derivation for
the PNC TPE $N\Delta$ potential \cite{delcon}, which
however is not utilized in this work.

The remainder of the paper is outlined as follows. Section
II gives the basic formalism for the calculation of the
PNC $\vec{p}p$ elastic scattering and Sec. III summarizes
the results.
\section{Formalism}\label{formalism}
The PNC $\vec{p}p$ elastic scattering experiments measure
the difference between the cross-sections $\sigma_{m_1}$
of the transmitted protons with the spins parallel
($m_1=\frac{1}{2}$) and antiparallel ($m_1=-\frac{1}{2}$)
along the direction of propagation. The PNC analyzing power
is given as
\begin{equation}
\bar{A}_L=
\frac{\sigma_\frac{1}{2}
-\sigma_{-\frac{1}{2}}}
{\sigma_\frac{1}{2}
+\sigma_{-\frac{1}{2}}},
\end{equation}
where in the other words $\sigma_\frac{1}{2}$ and
$\sigma_{-\frac{1}{2}}$ denotes respectively the total
cross-sections of the positively and negatively polarized
proton beam.

The complete potential for the PNC $pp$ interaction
is a combination of distortive PC potentials and
small perturbatively treated PNC potentials
$\hat{V}=
\hat{V}^{{\rm PC}}+
\hat{V}^{{\rm PNC}}$.
The PC potential is the sum of the Coulomb
$\hat{V}^{{\rm PC}}_{\rm C}$ and nuclear
$\hat{V}^{{\rm PC}}_{\rm N}$ $NN$ potentials,
where we take $\hat{V}^{{\rm PC}}_{\rm N}$ as the
Reid93 potential \cite{reid93}. The PNC potential is
considered to arise from the long-ranged TPE potential and
short-ranged heavy meson potential
$\hat{V}^{{\rm PNC}}=
\hat{V}^{{\rm PNC}}_{2\pi}+
\hat{V}^{{\rm PNC}}_{\rho,\omega}$.
%
%
%
The used PNC TPE potentials \cite{our1,desplanq,kaiser}
are respectively abbreviated by the authors as NPI, DHAL,
and K. The potentials $\hat{V}^{\rm DHAL}_{2\pi}$ and
$\hat{V}^{\rm K}_{2\pi}$ are built on QCD based ChPT
and the $\hat{V}^{\rm NPI}_{2\pi}$ on the time-ordered
perturbation theory. The $\hat{V}^{\rm DHAL}_{2\pi}$
results essentially from the $v_{44}^{\rm EFT}(q)$ in
Eq. 12 of Ref. \cite{desplanq}. Notable is that it
comprises only the $NN$ intermediate states while the
$\hat{V}^{\rm K}_{2\pi}$ and $\hat{V}^{\rm NPI}_{2\pi}$
include also the $N\Delta$ intermediate states.
Anyhow, they all are the spin changing local potentials
of the form $\hat{V}^{{\rm PNC}}_{2\pi}(\bm{r})
=h_\pi^{(1)}(\hat{\tau}_{1z}+\hat{\tau}_{2z})
(\bm{\sigma}_1\times\bm{\sigma}_2)\cdot\hat{\bm{r}}W(r)$
in the two-proton case. The radial functions $W(r)$
have different structures in each potential.
The $NN$ parts of the unregularized
$\hat{V}^{\rm K}_{2\pi}(\bm{r})$ and
$\hat{V}^{\rm DHAL}_{2\pi}(\bm{r})$ potentials are
identical, apart from the $\delta(\bm{r})$-function
term in the latter one arising from the constant term
in momentum space in its dispersion relation.
However, since we are not only dealing with low energies,
these two potentials should be provided with form factors,
in which case they differ from each other even if
regularized by the same form factors. The DDH "best" value
$h_\pi^{(1)} = 4.6\times 10^{-7}$ is used in the PNC TPE
potentials.

As a PNC heavy meson potential, we use the DDH potential
and their "best" weak meson-$NN$ coupling values. The
isospin matrix element of the DDH potential, taken between
the intial and final $pp$ states, is
%
\begin{align}\label{ppddh}
&\hat{V}^{{\rm PNC}}_{\rho,\omega}(\bm{r})=
-\sum_{\alpha=\rho,\omega}
\frac{g_\alpha h_\alpha^{pp}}{2M}
\Bigl((\bm{\sigma}_1-\bm{\sigma}_2)
\cdot\{-i\bm{\nabla},Y_\alpha(r)\}
+i(1+\chi_\alpha)
(\bm{\sigma}_1\times\bm{\sigma}_2)
\cdot[-i\bm{\nabla},Y_\alpha(r)]
\Bigr),
\end{align}
with $h_\rho^{pp}=
h_\rho^{(0)}+h_\rho^{(1)}+
h_\rho^{(2)}/\sqrt{6}$ and
$h_\omega^{pp}=
h_\omega^{(0)}+h_\omega^{(1)}$,
which have the numerical values of
$-15.48$ and $-3.00$ in units of $10^{-7}$
respectively. As for the other parameters,
we take the values for the strong couplings as
$g_\rho=2.79$ and  $g_\omega=8.37$ and for the
anomalies as $\chi_\rho=3.71$ and  $\chi_\omega=-0.12$.
The radial $Y_\alpha(r)=\exp(-m_\alpha r)/4\pi r$
are the Yukawa functions, which we use only in the
form modified by the dipole form factors of the type
$(\Lambda_\alpha^2-m_\alpha^2)^2
(\bm{q}^2+\Lambda_\alpha^2)^{-2}$ taking the cut-off
masses as $\Lambda_\rho=1.3$ GeV and
$\Lambda_\omega=1.5$ GeV. The
$V^{\rm NPI}_{2\pi}$ potential is slightly
scaled to correspond to the strong $\pi NN$ coupling
value of $g_\pi=13.45$, which is in line with the
other couplings. The heavy meson masses are
$m_\rho=770$ MeV and $m_\omega=782$ MeV, and $M=939$
MeV is the average nucleon mass. We call the couplings
given above as the standard set, since they are the
typical choice in the PNC calculations.

The $pp$ scattering amplitude
$f(k,\theta)=f^{\rm C}(k,\theta)+f^{\rm N}(k,\theta)$
consists of the Coulomb scattering amplitude
(superscripted by C) representing electromagnetic
interaction and the Coulomb-nuclear scattering amplitude
(superscripted by N) including electromagnetic, strong, and
weak interactions. The Coulomb scattering amplitude is given
by

\begin{equation}\label{coulamp}
f^{\rm C}(k,\theta)=
-\frac{\eta}{2k\sin^2\frac{\theta}{2}}
e^{i[2\sigma_0-\eta\ln\sin^2\frac{\theta}{2}]},
\end{equation}
where $\eta=\alpha\mu/k$, $\alpha$ is the fine-structure
constant, $\mu = M/2$ is the reduced mass of the two
nucleons, and $\sigma_0=\arg\Gamma(1+i\eta)$ is the
Coulomb S-wave phase shift. An awkward feature of Eq.
\eqref{coulamp} is that it is undefined at $\theta=0$.
Thus, in the determination of the total scattering
cross-section by means of the optical theorem, the
singularity of the total scattering amplitude in the
forward direction is simply removed by the subtraction
of the $f^{\rm C}(k,0)$, leaving only the $f^{\rm N}(k,0)$
to contribute. The forward, $\theta=0$, $\vec{p}p$
Coulomb-nuclear scattering amplitude in the DWBA is given
by
\begin{align}
f_{\substack{m_1m_2\\m_1m_2}}^{\rm N}(k,0)=
-\frac{\mu}{2\pi}
\Bigl[
{}_{\rm C}\langle k\hat{z};m_1m_2|
\hat{V}^{{\rm PC}}_{\rm N}|
k\hat{z};m_1m_2\rangle^{(+)}
+{}^{(-)}\langle k\hat{z};m_1m_2|
\hat{V}^{{\rm PNC}}|
k\hat{z};m_1m_2\rangle^{(+)}
\Bigr],
\end{align}
where the nuclear potentials are sandwiched between the
Coulomb-distorted strong interaction wavefunctions. The
$pp$ wavefunctions are of the form
\begin{align}\label{pnwfmm}
\langle\bm{r}|k\hat{z};m_1m_2\rangle^{(\pm)}=~&
\sum_{SM_S}
\langle{\textstyle\frac{1}{2}}m_1
{\textstyle\frac{1}{2}}m_2\vert SM_S\rangle
\langle\bm{r}|k\hat{z};SM_S\rangle^{(\pm)},
\end{align}
with
\begin{align}\label{pnwf}
\langle\bm{r}|k\hat{z};SM_S\rangle^{(\pm)}=~&
\frac{\sqrt{8\pi}}{kr}
\sum_{L'LJ}i^{L}\sqrt{2L+1}
\langle L0SM_S\vert JM_S\rangle e^{\pm i\sigma_L}
\mathcal{U}_{LL'}^{SJ(\pm)}(k,r)
\mathscr{Y}^{L'S}_{JM_S}(\hat{\bm{r}})
\vert 11\rangle,
\end{align}
where the z-axis is taken along the direction of $\bm{k}$,
$\mathscr{Y}^{L'S}_{JM_S}(\hat{\bm{r}})$ are the
eigenfunctions of the coupled angular momentum, and
$\vert 11\rangle$ denotes the isospin state
$\vert TM_T\rangle$ of the two-protons. In the wavefunctions
with the subscript C, as it is in final state of the PC
amplitude, the radial wavefunctions
$\mathcal{U}_{LL'}^{SJ(\pm)}(k,r)$ (including the phase
shifts) are simply replaced by the regular Coulomb functions
$F_L(kr)$, which further reduce to the spherical Bessel
functions $j_L(kr)$ if the Coulomb interaction is turned off,
$\it{i.e.}~\eta=0$. From the given equations, the longitudinal
scattering asymmetry becomes
\begin{align}\label{scatassym}
\bar{A}_L(k)
&=\frac{{\rm Im}
\sum_{SS'}{}^{(-)}\langle k\hat{z};S'0|
\hat{V}^{{\rm PNC}}|
k\hat{z};S0\rangle^{(+)}}{{\rm Im}
\sum_{SM_S}{}_{\rm C}\langle k\hat{z};SM_S|
\hat{V}^{{\rm PC}}_{\rm N}|
k\hat{z};SM_S\rangle^{(+)}}.
\end{align}

While the lower energy experiments measure directly
the scattered particles, the TRIUMF E497 and higher energy
experiments measure the transmitted beam after passing
through the target, see {\it e.g.} Ref. \cite{ramsay} for
a summary of the PNC $\vec{p}p$ experiments. In
transmission experiments, a complication arises due to the
fact the Coulomb interaction is singular in the forward
direction. Therefore, we consider the Coulomb distortions
near the propagation direction of the transmitted beam, as
done, {\it e.g.} in Refs. \cite{driscoll} and
\cite{carlson}.
Symmetrized and properly normalized Coulomb scattering
amplitude may be written as
\begin{equation}\label{expamp2}
f^{\rm C}_{\substack{m_1m_2\\m_1'm_2'}}(k,\theta)
=\frac{1}{\sqrt{2}}\sum_{SM_S}
\langle{\textstyle\frac{1}{2}}m_1
{\textstyle\frac{1}{2}}m_2|SM_S\rangle
\langle{\textstyle\frac{1}{2}}m_1'
{\textstyle\frac{1}{2}}m_2'|SM_S\rangle
\Bigl[f^{\rm C}(k,\theta)+
(-)^Sf^{\rm C}(k,\pi-\theta)\Bigr],
\end{equation}
where $f^{\rm C}(k,\theta)$ is given in Eq. \eqref{coulamp}.
The spin averaged Coulomb cross-section for a transmission
experiment becomes
\begin{equation}
\sigma_{m_1}^{\rm C_{\theta_0}}(k)=
\pi\sum_{\substack{m_2\\m_1'm_2'}}
\int_{\theta_0}^{\frac{\pi}{2}}
d\theta\sin\theta
|f_{\substack{m_1m_2\\m_1'm_2'}}
^{\rm C}(k,\theta)|^2
=\frac{\pi\eta^2}{2k^2}
\Bigl(\frac{1}{\sin^2\frac{\theta_0}{2}}
-\frac{1}{\cos^2\frac{\theta_0}{2}}
+\frac{1}{\eta}
\sin[2\eta\ln\tan\frac{\theta_0}{2}]\Bigr)
\delta_{m_1m_1},
\end{equation}
where $\theta_0>0$ is such a small cut-off angle
that $f^{\rm N}(k,\theta_0)\approx f^{\rm N}(k,0)$.
The corresponding nuclear cross-section is
\begin{equation}\label{nucc}
\sigma_{m_1}^{\rm N_{\theta_0}}(k)=
\sigma_{m_1}^{\rm N}(k)
-2\pi\int_0^{\theta_0}
d\theta\sin\theta
\frac{d\sigma_{m_1}^{\rm N}}{d\Omega}(k,\theta)
=\frac{\pi}{k}\sum_{m_2}{\rm Im}
\Bigl(f_{\substack{m_1m_2\\m_1m_2}}^{\rm N}(k,0)
e^{2i[\eta\ln\sin\frac{\theta_0}{2}-\sigma_0]}
\Bigr),
\end{equation}
where $\sigma_{m_1}^{\rm N}(k)$ is the total
cross-section given by the optical theorem
and the differential cross-section is taken as
$d\sigma_{m_1}^{\rm N}
=d\sigma_{m_1}-d\sigma_{m_1}^{\rm C}.$
In the last step of Eq. \eqref{nucc}, the result
\begin{equation}
\int_{\epsilon\rightarrow 0}^{\theta_0}
d\theta\sin\theta
f^{{\rm C}\ast}(k,\theta)=
\frac{1}{ik}\Bigl(
1-e^{2i[\eta\ln\sin\frac{\theta_0}{2}-\sigma_0]}
\Bigr),
\end{equation}
first derived in Ref. \cite{hold}, was used.
The longitudinal transmission asymmetry becomes
\begin{align}\label{transassym}
\bar{A}^{\theta_0}_L(k)&=
\frac{{\rm Im}\Bigl[
\sum_{SS'}{}^{(-)}\langle k\hat{z};S'0|
\hat{V}^{{\rm PNC}}|
k\hat{z};S0\rangle^{(+)}
e^{2i[\eta\ln\sin\frac{\theta_0}{2}-\sigma_0]}
\Bigr]}{{\rm Im}\Bigl[
\sum_{SM_S}{}_{\rm C}\langle k\hat{z};SM_S|
\hat{V}^{{\rm PC}}_{\rm N}|
k\hat{z};SM_S\rangle^{(+)}
e^{2i[\eta\ln\sin\frac{\theta_0}{2}-\sigma_0]}
\Bigr]-\frac{4k}{M}
\sigma^{\rm C_{\theta_0}}(k)}.
\end{align}
\section{Results}

\begin{figure}[b]
\includegraphics[scale=1.2]{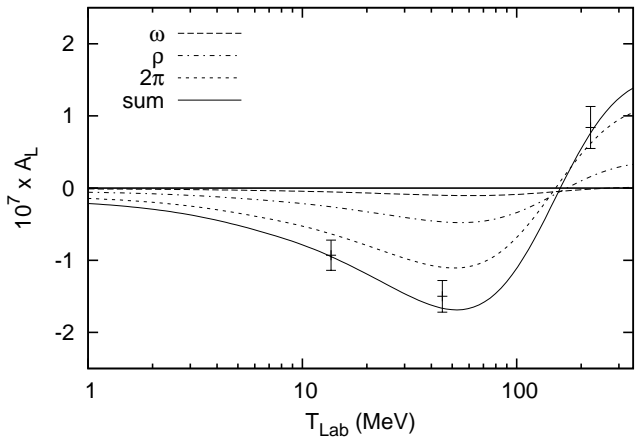}
\caption{\label{scatasym} The
contributions of the $\rho$-, $\omega$-, and $2\pi$-
exchanges to the analyzing power.}
\end{figure}
\begin{figure}[t]
\includegraphics[scale=1.2]{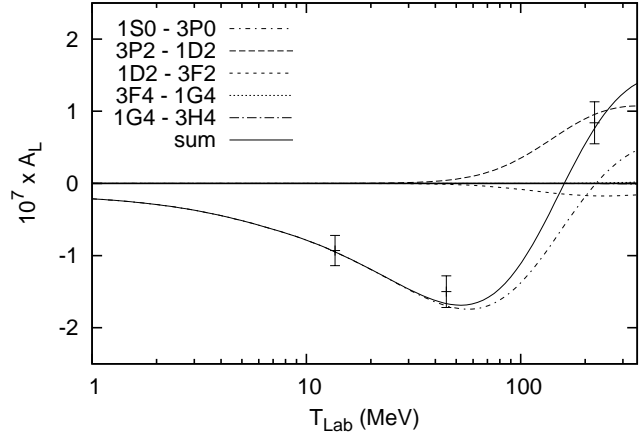}
\caption{\label{partial} The
partial-wave contributions of the total
scattering asymmetry.}
\end{figure}
\begin{figure}[b]
\includegraphics[scale=1.2]{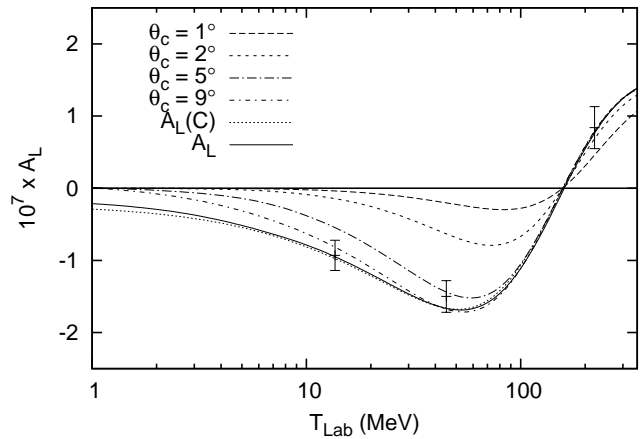}
\caption{\label{critical} The scattering asymmetry of Eq.
\eqref{scatassym} with ($A_L(\rm{C})$) and without
($A_L$) long-range Coulomb effects and the different
cut-off angle $\theta_{\rm c}$ transmission asymmetries
of Eq. \eqref{transassym} are illustrated.}
\end{figure}
\begin{figure}[t]
\includegraphics[scale=1.2]{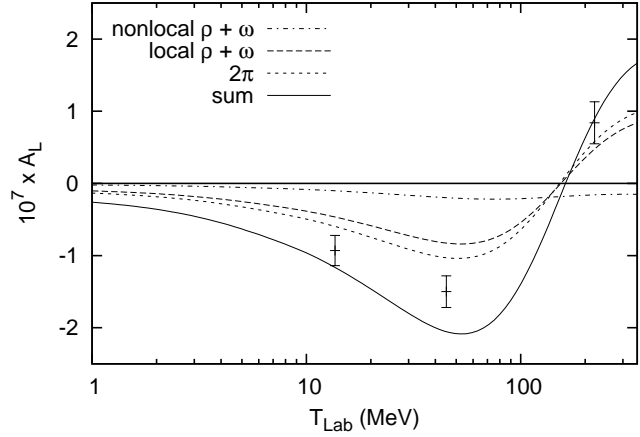}
\caption{\label{bonncoup} Different aspects of the scattering asymmetry given by the strong couplings related
to the Bonn potential.}
\end{figure}

Now we give the results at energies ranging between 1
and 350 MeV for the PNC longitudinal analyzing powers.
In all cases the Reid93 potential is employed and the
shown experimental data points are the Bonn, PSI, and TRIUMF
ones, for which the values are given in the introduction.
In Figs. \ref{scatasym}-\ref{critical} we use the standard
set of couplings given in Sec. \ref{formalism} whereas in
Fig. \ref{bonncoup} we use a weaker $\pi NN$ coupling
together with the heavy meson couplings of the configuration
space Bonn potential \cite{mach}. In all Figs.
\ref{scatasym}-\ref{bonncoup}, we employ the
$\hat{V}^{\rm NPI}_{2\pi}$ as a reference TPE potential
along with the PNC heavy meson exchange potential Eq.
\eqref{ppddh} with the dipole form factors of the type
$(\Lambda_\alpha^2-m_\alpha^2)^2 (\bm{q}^2+\Lambda_\alpha^2)^{-2}$ with the cut-offs of
$\Lambda_\rho=1.3$ GeV and $\Lambda_\omega=1.5$ GeV.
Basically, the PNC $pp$ effects are exclusive properties
of nuclear interactions disturbed by the Coulomb field
within the range of the nuclear forces. To obtain a
clean PNC signal, the external long-range Coulomb effects
can be cut out. In all figures, we utilize the scattering
analyzing power of Eq. \eqref{scatassym}, in which the
long-range Coulomb effects are neglected by omitting the
Coulomb phases $e^{i\sigma_L}$ of the wavefunctions in Eq.
\eqref{pnwf}. However, these negligible effects are
included in the asymmetries of Fig. \ref{critical},
where also the transmission analyzing power of Eq.
\eqref{transassym} is depicted.

%
Figure \ref{scatasym} shows separately the $\rho$-,
$\omega$-, and TPE contributions to the asymmetry.
Throughout the energy range, the TPE effect is about twice
as large as the that of the heavy mesons and, as a
consequence, the calculated asymmetry sets within the error
limits of the experimental data.
Figure \ref{partial} depicts the contributions of the
different parity admixed partial waves up to $J=4$. The
transitions with $J=4$ (or higher) are unimportant and,
thus, the lowest three admixtures would in fact be
sufficient within the used energy range. One particularly
interesting feature of the asymmetry, as was first pointed
out in Ref. \cite{simonius} and utilized in the TRIUMF
experiment, is that the ${}^1S_0-{}^3P_0$ contribution
vanishes at a specific energy due to the equal, but
opposite phase shifts of the ${}^1S_0$ and ${}^3P_0$
partial waves, which is seen at 224.7 MeV in Fig.
\ref{partial}.

The long-range Coulomb effects to the asymmetries are
illustrated in Fig. \ref{critical} along with the cut-off
angle $\theta_{\rm c}$ dependence of the transmission
asymmetry.
The calculated scattering and transmission asymmetries at
the energies of about 150 MeV and above become nearly
indistinguishable by the angles $\theta_{\rm c}\geq2^\circ$.
Especially noteworthy is that at the energy of the only
transmission experiment, TRIUMF, the asymmetry remains practically unaffected.
Just to show the strong coupling sensitivity to the
analyzing power, in Fig. \ref{bonncoup} we employ an
alternative set of couplings, $g_\pi^2/4\pi=13.8$ for the
pion and the Bonn potential \cite{mach} configuration space
values $g_\rho^2/4\pi=0.95$, $g_\omega^2/4\pi=20$,
$\chi_\rho=6.1$, and $\chi_\omega=0$ for the heavy mesons.
Compared to the use of the standard set of couplings, the
asymmetry is enhanced by this choice of couplings. Also the
TPE and heavy meson exchange contributions to the analyzing
power become about equal. In the same figure, we have
separated the nonlocal and local contributions of the PNC
heavy meson exchange potential, which arise respectively
from the anticommutator and commutator terms of Eq.
\eqref{ppddh}. The resulting curves are formally consistent
with Ref. \cite{driscoll}. Note that the scaling between
Figs. \ref{scatasym} and \ref{bonncoup} is straightforward
for $\rho+\omega$ total and TPE, since only the
aforementioned strong couplings are changed.

\begin{figure}[t]
\includegraphics[scale=1.2]{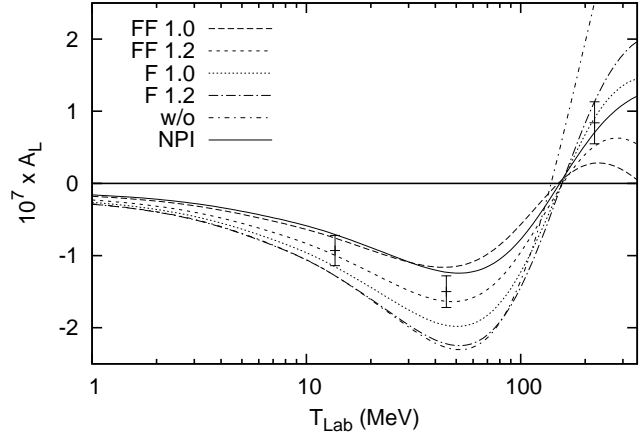}
\caption{\label{kai2pi} The TPE by the
$\hat{V}^{\rm K}_{2\pi}(\bm{r})$ potential with the monopole
(F) and dipole (FF) form factors using the cut-off masses
$\Lambda=1.0$ GeV and 1.2 GeV and also without (w/o) the
form factors. As a comparison, the asymmetry is also plotted
using the $\hat{V}^{\rm NPI}_{2\pi}(\bm{r})$ potential in
which the coupling values are scaled to correspond to those
of the $\hat{V}^{\rm K}_{2\pi}(\bm{r})$.}
\end{figure}

\begin{figure}[t]
\includegraphics[scale=1.2]{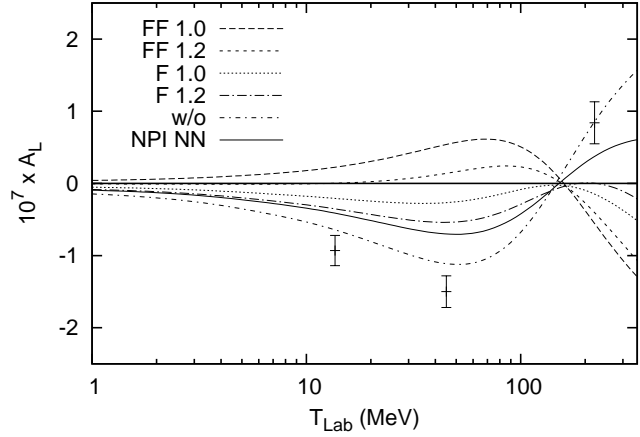}
\caption{\label{des2pi} The same as Fig. \ref{kai2pi} but
the TPE is by the $\hat{V}^{\rm DHAL}_{2\pi}(\bm{r})$
potential and the $NN$ part of the
$\hat{V}^{\rm NPI}_{2\pi}(\bm{r})$ potential. The "w/o"
curve is identical with that of the $NN$ part of the
$\hat{V}^{\rm K}_{2\pi}(\bm{r})$.}
\end{figure}

Figures \ref{kai2pi} and \ref{des2pi} represent
respectively the PNC TPE potential $\hat{V}^{\rm K}_{2\pi}$
and $\hat{V}^{\rm DHAL}_{2\pi}$ contributions to the
scattering asymmetry. Because the PNC potentials are
in general treated perturbatively in the DWBA, the
regularization of them is not vital. However, since the
effect of the singularity comes forth more and more along
with the increasing energy, it should be removed by the
regularization as usual. In contrast, the chiral
perturbation theory based $\hat{V}^{\rm K}_{2\pi}$
and $\hat{V}^{\rm DHAL}_{2\pi}$ potentials would
serve their purpose best as unregularized due to their
model independent nature. However, in Figs. \ref{kai2pi}
and \ref{des2pi} these potentials are used both with
(F or FF) and without (w/o) regularization. When
regularized, we incorporate the monopole
$\Lambda^2(\bm{q}^2+\Lambda^2)^{-1}$ (F) and dipole
$\Lambda^4(\bm{q}^2+\Lambda^2)^{-2}$ (FF) form factors
using two different cut-off masses $\Lambda=1.0$ GeV and
$\Lambda=1.2$ GeV.
A monopole form factor of the same type
is also used to the $np$ part of the
$\hat{V}^{\rm DHAL}_{2\pi}$ potential in the
radiative reaction $\vec{n}p\rightarrow \gamma d$ in Ref.
\cite{hyun2pi}. As seen in Figs. \ref{kai2pi} and
\ref{des2pi}, the resulting asymmetries are in most cases
formally similar for all the
$\hat{V}^{\rm K}_{2\pi}$,
$\hat{V}^{\rm DHAL}_{2\pi}$, and
$\hat{V}^{\rm NPI}_{2\pi}$ potentials.
When switching over from the monopole to dipole type form
factor and from larger cut-off to smaller, the diminishing
effect on the TPE becomes stronger.
Figure \ref{kai2pi} shows that when using the dipole form
factor with $\Lambda=1.0$ GeV, the effect of the
$\hat{V}^{\rm K}_{2\pi}$ up to about 150 MeV is more or
less indistinguishable from the one of the
$\hat{V}^{\rm NPI}_{2\pi}$. In other cases, the asymmetry
is larger.
As illustrated in Fig. \ref{des2pi}, the asymmetry using the
$\hat{V}^{\rm DHAL}_{2\pi}$ is very sensitive to the used
regularizations, because of the form factor modified
$\delta$-term, and even exhibits a different sign when the
dipole form factor is used. The asymmetry is also chiefly
smaller than the reference (NPI $NN$) curve if the
regularization is used. In unregularized form, the
$\hat{V}^{\rm DHAL}_{2\pi}$ and the $NN$ part of the
$\hat{V}^{\rm K}_{2\pi}$ coincide from which follows that
the "w/o" curve in Fig. \ref{des2pi} is identical for each
one of these two potentials. This curve is larger
than the NPI one, but becomes roughly the same in the case
of the $\hat{V}^{\rm K}_{2\pi}$ with dipole form factor and
$\Lambda=1.0$ GeV. Bringing up the difference between the
$NN$ and $NN+N\Delta$ for the $\hat{V}^{\rm K}_{2\pi}$
and $\hat{V}^{\rm NPI}_{2\pi}$ compare with Figs.
\ref{kai2pi} and \ref{scatasym}
(or \ref{partial}-\ref{critical}) respectively.
Lastly, minimizing the model dependence of the TPE,
one may speculate on the value of the $h_\pi^{(1)}$.
Some heuristic estimates of it can be "eyeballed" off Figs.
\ref{kai2pi} and \ref{des2pi} by considering only the "w/o"
curves and assuming that the heavy meson effect of Fig.
\ref{scatasym} represents realistically the short-range
contribution to the analyzing power.
The $\hat{V}^{\rm K}_{2\pi}$ suggests that the value of the
$h_\pi^{(1)}$ should be roughly $50\%$ smaller while
$\hat{V}^{\rm DHAL}_{2\pi}$ and $NN$ part of the
$\hat{V}^{\rm K}_{2\pi}$ that the "DDH" best value is about
correct.

In summary, we have calculated the PNC longitudinal
analyzing power $\bar{A}_L(\vec{p}p\rightarrow pp)$ by
taking into account the electromagnetic and TPE effects in
various models. Coulomb interaction plays virtually no role
in the scattering or transmission asymmetries. By using the
aforementioned standard set of couplings, we found that the
$\hat{V}^{\rm NPI}_{2\pi}$ potential along with the DDH model
gives an excellent match with the experimental data. The TPE
effect is about two times larger than heavy meson exchange
effect throughout the energy scale.
Nearly consistent result comes also from the
$\hat{V}^{\rm K}_{2\pi}$ potential with the dipole form
factor and $\Lambda=1.0$ GeV cut-off. The above model is also used in the calculation of the cold neutron spin rotation
$\frac{d}{dz}\phi$, polarization $\frac{d}{dz}P$, and
$\gamma$-asymmetry
$\mathcal{A}_\gamma(\vec{n}p\rightarrow\gamma d)$ in the
interaction with parahydrogen \cite{partanen}. Within the
said model, the TPE effect reasonably diminishes the OPE
effect by about $10\%$ in the observables.
%
%
All in all, assuming that the two-pion and heavy meson
exchanges are the only major contributions to the analyzing
power $\bar{A}_L$, we found that the $\bar{A}_L$ depends
mostly on the TPE unless the true value of $h_\pi^{(1)}$ is
significantly smaller than that given by DDH. However,
ultimately, the experiments ({\it e.g.} the NPDGamma
experiment) may decide the reliability of this value.
As a conclusion of this work, despite the inescapable model
dependence of the observable $\bar{A}_L$, the TPE causes
most likely an important effect to it and should not be
ignored.
%
%
%
%
\begin{acknowledgements}
T. M. P. is grateful for the hospitality of the University
of British Columbia and would like to thank Dr. M.~J.~Iqbal
for hosting the stay at the UBC. T. M. P. also gratefully
acknowledges the financial support from the Vilho,
Yrj\"{o}, and Kalle V\"{a}is\"{a}l\"{a} Foundation.
This work was partly supported by an Academy of Finland
researcher exchange grant 139512.
\end{acknowledgements}
%
%
%
\newpage
%
%
%

\end{document}